\documentclass[sigconf]{acmart}

\AtBeginDocument{%
  }

\usepackage{multirow}
\usepackage{soul}
\usepackage{enumitem}
\usepackage{xcolor}
\usepackage{bm}
\usepackage{framed}

\copyrightyear{2025}
\acmYear{2025}
\setcopyright{acmlicensed}
\acmConference[SIGIR '25]{Proceedings of the 48th International ACM SIGIR Conference on Research and Development in Information Retrieval}{July 13--18, 2025}{Padua, Italy}
\acmBooktitle{Proceedings of the 48th International ACM SIGIR Conference on Research and Development in Information Retrieval (SIGIR '25), July 13--18, 2025, Padua, Italy}
\acmDOI{10.1145/3726302.3730019}
\acmISBN{979-8-4007-1592-1/2025/07}

\begin{document}

\title{Large Language Models Enhanced Hyperbolic Space Recommender Systems}

\author{Wentao Cheng}
\orcid{0009-0001-0698-5838}
\affiliation{%
  \institution{Beijing Institute of Technology}
  \city{Beijing}
  \country{China}}
\additionalaffiliation{%
  \institution{Guangdong Laboratory of Artificial Intelligence and Digital Economy (SZ)}
  \city{Shenzhen}
  \country{China}}
\email{wentao.cheng23@gmail.com}

\author{Zhida Qin}
\orcid{0000-0002-9270-1810}
\authornote{Corresponding authors.\\Our code is released at \url{https://github.com/Qin-lab-code/HyperLLM}.
}
\affiliation{%
  \institution{Beijing Institute of Technology}
  \city{Beijing}
  \country{China}
}
\email{zanderqin@bit.edu.cn}

\author{Zexue Wu}
\orcid{0009-0001-6953-4911}
\affiliation{%
  \institution{Beijing Institute of Technology}
  \city{Beijing}
  \country{China}}
\email{zexue.wu@bit.edu.cn}

\author{Pengzhan Zhou}
\orcid{0000-0002-8796-5969}
\affiliation{%
  \institution{Chongqing University}
  \city{Chongqing}
  \country{China}}
\email{pzzhou@cqu.edu.cn}

\author{Tianyu Huang}
\orcid{0000-0003-2307-2261}
\affiliation{%
  \institution{Beijing Institute of Technology}
  \city{Beijing}
  \country{China}}
\email{huangtianyu@bit.edu.cn}

\renewcommand{\shortauthors}{Cheng et al.}

\begin{abstract}
Large Language Models (LLMs) have attracted significant attention in recommender systems for their excellent world knowledge capabilities. However, existing methods that rely on Euclidean space struggle to capture the rich hierarchical information inherent in textual and semantic data, which is essential for capturing user preferences. The geometric properties of hyperbolic space offer a promising solution to address this issue. Nevertheless, integrating LLMs-based methods with hyperbolic space to effectively extract and incorporate diverse hierarchical information is non-trivial. To this end, we propose a model-agnostic framework, named \textbf{HyperLLM}, which extracts and integrates hierarchical information from both structural and semantic perspectives. Structurally, HyperLLM uses LLMs to generate multi-level classification tags with hierarchical parent-child relationships for each item. Then, tag-item and user-item interactions are jointly learned and aligned through contrastive learning, thereby providing the model with clear hierarchical information. Semantically, HyperLLM introduces a novel meta-optimized strategy to extract hierarchical information from semantic embeddings and bridge the gap between the semantic and collaborative spaces for seamless integration. Extensive experiments show that HyperLLM significantly outperforms recommender systems based on hyperbolic space and LLMs, achieving performance improvements of over 40\%. Furthermore, HyperLLM not only improves recommender performance but also enhances training stability, highlighting the critical role of hierarchical information in recommender systems.
\end{abstract}



\maketitle

\section{Introduction}
Recommender systems have become fundamental to various industries, from e-commerce to content streaming. Traditional recommender techniques ~\cite{karatzoglou2017deep, he2020lightgcn} are effective, but they often face challenges in handling complex relationships between users and items, especially when dealing with large-scale and sparse data. As the demand for personalized recommendations increases, recent advances ~\cite{zhang2022geometric, xu2023geometry} in geometric methods provide promising solutions to address these challenges and improve performance.

Hyperbolic geometry is a non-Euclidean space that has gained increasing attention for its ability to model hierarchical structures and complex relationships ~\cite{sala2018representation, ganea2018hyperbolic, chami2019hyperbolic}. Unlike the Euclidean space, where spatial capacity grows linearly, hyperbolic space exhibits exponential growth, making it particularly suitable for representing hierarchical data, such as taxonomies and social networks. Research ~\cite{zhang2021we, sun2021hgcf, wang2021fully} indicates that hierarchical structures exist between users and items in recommender systems. The unique properties of hyperbolic space enable it to effectively model these relationships, thereby better capturing user preferences. 

In recent years, with the widespread application of Large Language Models (LLMs) ~\cite{kenton2019bert, brown2020language} across various fields, their potential to enhance recommender systems has attracted significant attention. A major category of LLMs-based recommender systems ~\cite{ren2024enhancing, liu2024collaborative, wang2024large} focuses on utilizing the excellent world knowledge capabilities of LLMs to extract and enrich semantic information from recommender data, addressing challenges such as data sparsity and long-tail effect. However, recent studies ~\cite{chen2024hyperbolic, yang2024enhancing} show that semantic information often contains rich hierarchical structures, and methods relying on Euclidean space may struggle to capture these structures, resulting in performance degradation. The hyperbolic space offers a promising direction to address this issue.

However, it is non-trivial to integrate LLMs-based methods with hyperbolic space, which presents challenges arising from two aspects. The first challenge lies in how to extract the diverse hierarchical information hidden within textual and semantic data. The second challenge is the significant gap between the semantic and collaborative information, which makes it difficult to incorporate the extracted hierarchical information into recommender systems. Moreover, this gap is further exacerbated by the differences between hyperbolic and Euclidean spaces.
\textbf{To address these challenges, we extract and integrate hierarchical information from structural and semantic perspectives.}

From the structural perspective, we focus on well-defined hierarchical relationships in the data. Specifically, we use LLMs to generate multi-level tags with parent-child relationships for each item, representing classifications ranging from broad to specific. These tags are inspired by classification networks ~\cite{horta2023semantic} in real-world taxonomies, which exhibit clear hierarchical structures. To incorporate this information into the hyperbolic model, we construct a tag-item matrix and jointly learn the tag-item and user-item interactions. Subsequently, hyperbolic space contrastive learning ~\cite{qin2024hyperbolic} aligns item representations from both sides, collectively providing structural hierarchical information to the model.

Furthermore, we extract semantic hierarchical information to capture more nuanced aspects of the data. Generally, the original textual data describes the objective characteristics of items but rarely reflects subjective user preferences, resulting in a lot of noise unrelated to preferences. Therefore, we use LLMs to summarize preference related information for both users and items. These summaries are converted into semantic embeddings using a text embedding model ~\cite{neelakantan2022text}. 
Then, we propose a novel meta-optimized two-phase training strategy, where each phase corresponds to a distinct training objective. In the first phase, the primary objective is to extract semantic hierarchical information from the semantic embeddings. During this phase, we freeze the parameters of both the semantic embeddings and the hyperbolic model, training only the Mixture of Experts (MoE) model ~\cite{jacobs1991adaptive, shazeer2017, gale2023megablocks} to transform the semantic embeddings effectively. In the second phase, we integrate the extracted hierarchical information into the hyperbolic model for recommendations. For meta-learning methods ~\cite{qin2023meta, hao2024meta}, freezing parameters facilitates the model to learn general knowledge from the target task, which improves generalization on new tasks. To distinguish between these two phases and encourage the model to learn more generalizable knowledge, we design task-specific adaptive margin ranking losses for each phase. \textbf{Note that the structural and semantic approaches are model-agnostic.}

In conclusion, we use LLMs to extract structural and semantic hierarchical information and propose a model-agnostic framework, named \textbf{HyperLLM}, which effectively integrates this information into hyperbolic space recommender systems. Consequently, HyperLLM achieves significant performance improvements on multiple hyperbolic baselines, with the highest improvement exceeding 40\%.

Our contributions can be summarized as follows:

$\bullet$ We leverage LLMs to extract both structural and semantic hierarchical information, taking into account the nature of recommender data and the properties of hyperbolic space. This approach enhances the recommender system's capability to accurately capture hierarchical relationships and user preferences.

$\bullet$ We introduce a novel meta-optimized strategy for extracting semantic hierarchical information with hyperbolic model, bridging the gap between semantic and hyperbolic collaborative spaces. This approach also provides a promising direction for semantic fusion in both Euclidean and hyperbolic space recommender systems.

$\bullet$ We conduct extensive experiments demonstrating the benefits of structural and semantic hierarchical information. Our proposed HyperLLM framework outperforms existing LLMs-based recommender frameworks in both performance and training stability, highlighting the importance of hierarchical information.

\section{Related Work}
\subsection{Hyperbolic Space Recommender Systems}
The geometric properties of hyperbolic space make hyperbolic neural networks highly effective for modeling data with hierarchical structures. In recommender systems, this ability to capture the inherent hierarchical relationships in user-item interactions has been increasingly recognized. Early efforts in this field focus on adapting techniques from Euclidean space. 
For example, HyperML~\cite{vinh2020hyperml} and HME~\cite{feng2020hme} introduce metric learning into hyperbolic space recommender systems. 
HVAE~\cite{mirvakhabova2020performance} designs a variational autoencoder model on hyperbolic space. HGCF~\cite{sun2021hgcf} proposes a Hyperbolic Graph Convolution Network (HGCN). These studies demonstrate the advantages of hyperbolic space through notable performance. Consequently, recent research has explored more efficient ways to leverage hyperbolic space. For instance, HRCF~\cite{yang2022hrcf} proposes a hyperbolic geometric aware regularizer loss, and HICF~\cite{yang2022hicf} enhances margin ranking loss with adaptive margins. However, due to the challenges in formalizing neural operations within hyperbolic space, earlier methods often map vectors to Euclidean space for computation, then back to hyperbolic space, resulting in suboptimal utilization of the hyperbolic space. To address this limitation, LGCF~\cite{wang2021fully}, GGCF~\cite{xu2023geometry}, and HGCH~\cite{zhang2024hgch} respectively propose fully HGCN on Klein, Hyperboloid, and Poincaré hyperbolic spaces, which perform computation entirely within hyperbolic space. 

\subsection{LLMs-based Recommender Systems}
Existing LLMs-based recommender systems can be broadly divided into two categories. The first category utilizes LLMs directly as the recommender system by applying strategies such as pre-training or fine-tuning LLMs. Early work in this category, such as P5~\cite{geng2022recommendation}, models multiple recommender tasks as a unified pre-training framework. 
ZeroShot~\cite{hou2024large} applies zero-shot techniques to LLMs for predicting recommendations without the need for fine-tuning. 
Recent studies mainly focus on enhancing LLMs from multiple perspectives, including performance, efficiency, and others. 
CLLM4Rec~\cite{zhu2024collaborative} designs a polynomial prediction head that enables the LLMs to predict multiple items in one step, avoiding auto-regressive prediction. 
TLRec~\cite{lin2024tlrec} uses the chain-of-thought technique to help LLMs better understand recommender data.
LC-Rec~\cite{zheng2024adapting} employs vector quantization to map items into learnable vectors that are compatible with LLMs, enhancing the collaborative semantics in LLMs.

The second category enhances existing recommender systems through strategies such as generating additional textual data or semantic representations. LLMRec~\cite{wei2024llmrec} uses LLMs to augment interaction graph and textual attributes to address the problem of data sparsity. SAID~\cite{hu2024enhancing} employs LLMs to learn semantically aligned item embeddings, which are then used to enhance downstream models. LRD~\cite{yang2024sequential} leverages LLMs to extract latent relations between items. KAR~\cite{xi2024towards} utilizes LLMs to generate reasoning and factual knowledge, which are then transformed into vectors by a hybrid-expert adaptor, making them compatible with recommender systems.
Furthermore, some studies propose novel strategies to bridge the gap between semantic information and collaborative information.
RLMRec~\cite{ren2024representation} employs contrastive learning and masked learning techniques to align collaborative representations with semantic ones.
CARec~\cite{wang2024collaborative} uses MSE loss to align collaborative and semantic representations in two bidirectional stages.

\section{Preliminaries}
\textbf{Hyperbolic space} is a non-Euclidean space with constant negative curvature. It can be described using various mathematical models, including the Hyperboloid model, the Poincaré model, the Klein model, and others. These models differ in how they represent the distance, curvature, and symmetry of hyperbolic space. In recommender systems, the Hyperboloid and Poincaré models are commonly used due to their ability to effectively represent hierarchical structures and complex relationships. 

For the $d$-dimensional hyperbolic space with constant negative curvature $c$, the Hyperboloid model $\mathbb{H}^d$ is represented as:
\begin{equation}
    \mathbb{H}^d = \{ \mathbf{x} \in \mathbb{R}^{d+1} \mid \left\langle \mathbf{x}, \mathbf{x} \right\rangle_\mathcal{\mathbb{H}} = 1/c, x_0 > 0 \}
\end{equation}
where $\left\langle \cdot, \cdot \right\rangle_\mathcal{\mathbb{H}}$ denotes the inner product in Hyperboloid model:
{\begin{equation}
\left\langle \mathbf{x} , \mathbf{y} \right\rangle_\mathbb{H}=-x_0y_0+\sum_{i=1}^dx_iy_i.
\end{equation}}

In addition, the Poincaré model $\mathbb{P}^d$ is represented as: 
\begin{equation}
    \mathbb{P}^d = \left\{ \mathbf{x} \in \mathbb{R}^d \mid \| \mathbf{x} \| < -1/c \right\}
\end{equation}
where $\| \mathbf{x} \|$ is the Euclidean norm of the point $\mathbf{x}$.

In Euclidean space, the inner product is used to calculate the angle between vectors and measure similarity. 
However, in hyperbolic space, the inner product does not directly reflect similarity between points, so distance metrics are typically used instead.

The distance between two points $\mathbf{x}, \mathbf{y} \in \mathbb{H}^d$ is defined as:
\begin{equation}
	d_\mathcal{\mathbb{H}}\left( \mathbf{x},\mathbf{y} \right)=\sqrt{k} \cdot \rm{arcosh}\left( -\left\langle \mathbf{x} , \mathbf{y} \right\rangle_\mathcal{\mathbb{H}} / k \right)
\end{equation}
where $k=-1/c$ is the negative reciprocal of $c$ and the distance between two points $\mathbf{x}, \mathbf{y} \in \mathbb{P}^d$ is defined as:
\begin{equation}
d_{\mathbb{P}}\left(\mathbf{x}, \mathbf{y}\right) = \sqrt{k} \cdot \rm{arcosh} \left( 1 + \frac{2k \cdot \| \mathbf{x} - \mathbf{y} \|^2}{\left(k - \| \mathbf{x} \|^2\right)\left(k - \| \mathbf{y} \|^2\right)} \right).
\end{equation}

This property also makes margin ranking loss commonly used in hyperbolic space recommender systems, as it has been shown to be useful in distance-based methods. 

The \textbf{hyperbolic margin ranking loss} $\mathcal{L}_{m}$ is expressed as:
\begin{equation}
\mathcal{L}_{m} \left( u,i,j\right) = \max \left(d\left(\mathbf{h}_u, \mathbf{h}_i \right) - d\left( \mathbf{h}_u, \mathbf{h}_j \right) + m, 0\right),
\end{equation}
where $u$ is user, $i$ and $j$ are the positive and negative items of $u$; $\mathbf{h}_u$, $\mathbf{h}_i$, and $\mathbf{h}_j$ are the embeddings of the user and items; $d(\cdot, \cdot)$ is the distance metric (e.g., $d_\mathcal{\mathbb{H}}$, $d_\mathcal{\mathbb{H}}^2$); and $m$ is the margin hyper-parameter.

The margin $m$ defines the minimum separation required between the positive and negative item distances in order to minimize the loss. If the difference between the distances is less than $m$, the loss penalizes the model, encouraging it to learn embeddings that better separate relevant items from irrelevant ones. The goal is to ensure that the model learns to push positive items closer to the user’s embedding while pushing negative items farther away, ultimately improving its ability to make accurate prediction.

In \textbf{meta-learning}, freezing parameters is a fast adaptation technique that enables the model to generalize more effectively to new tasks by optimizing the target task. In our proposed meta-optimized two-phase strategy, the target task involves extracting semantic hierarchical information in the first phase, while the new task corresponds to the recommender task in the second phase. To differentiate between these two tasks while promoting shared knowledge across them, we apply hyperbolic margin ranking losses in both phases, with margins tailored to the characteristics of each task.

\begin{figure*}[t]
    \vspace{-0.2in}
    \centering
    \includegraphics[width=17.6cm]{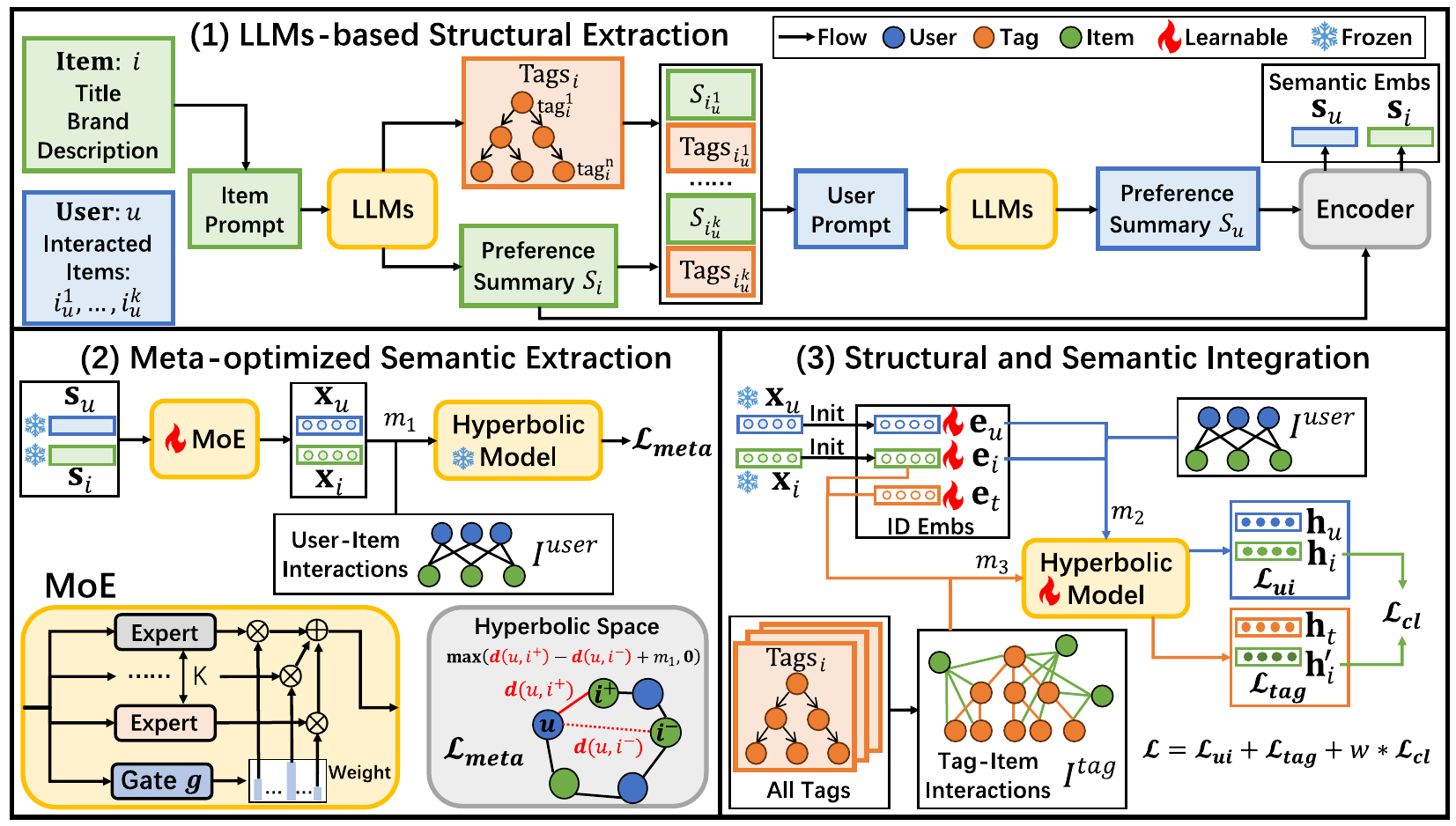}
    \captionsetup{skip=1pt}
    \caption{The overall architecture of our proposed HyperLLM. It consists of three modules: LLMs-based Structural Extraction, Meta-optimized Semantic Extraction, and Structural and Semantic Integration. These modules operate in the specified order.}
    \label{fig:HyperLLM}
    \vspace{-0.2in}
\end{figure*}

\section{Framework}
As illustrated in Figure \ref{fig:HyperLLM}, our proposed \textbf{HyperLLM} framework consists of three modules: \textbf{LLMs-based Structural Extraction}, \textbf{Meta-optimized Semantic Extraction}, and \textbf{Structural and Semantic Integration}.
The LLMs-based Structural Extraction module uses LLMs to extract structural hierarchical information and summarize preference information from recommender data. Building on this, the Meta-optimized Semantic Extraction module introduces a meta-optimized strategy to extract semantic hierarchical information from the embeddings of preference summaries. Finally, the Structural and Semantic Integration module integrates the structural and semantic hierarchical information into hyperbolic space recommender systems to construct a unified framework. 

In addition, the Meta-optimized Semantic Extraction module corresponds to \textbf{the first phase} of the meta-optimized two-phase training strategy, while the Structural and Semantic Integration module corresponds to \textbf{the second phase}. \textbf{The training of these two phases is decoupled, indicating that the second phase begins only after the first phase is completed.}

\subsection{LLMs-based Structural Extraction} \label{sec:hierarchicalinformationextraction}
In textual-based recommender systems, the data typically consists of user-item interactions and associated textual data, such as item titles, brands, descriptions, and other relevant metadata. We assume the user-item interactions to be a matrix $\mathcal{I}^{user}$, where each entry $\mathcal{I}_{ui}^{user}$ indicates the interaction between user $u$ and item $i$. Additionally, the textual data is represented as $\mathcal{T}_i = \{ t_i^1, t_i^2, \dots, t_i^m \}$, where $t_i^j$ refers to the $j$-th textual attribute of item $i$. 

\textbf{To extract structural hierarchical information}, we use LLMs to \textbf{generate multi-level tags with parent-child relationships} for each item. These tags represent broad-to-specific classifications, and the hierarchy of the tags is defined by their relationships, reflecting clear hierarchical structures similar to real-world taxonomies. 

Specifically, for each item $i$, we generate a set of tags as $\text{Tags}_i = \{ \text{tag}_{i}^1, \dots, \text{tag}_{i}^{n}\}$, where $\text{tag}_i^j$ is the $j$-th tag, ranging from general categories to more specific ones. 
Additionally, we generate edges that define the parent-child relationships between these tags, where higher-level tags act as parents and their related lower-level tags act as children. These edges are established for all tags generated by the LLMs. Formally, we define a set of edges $\mathcal{E} = \{ ( \text{tag}^a, \text{tag}^b ) \mid \text{tag}^a \to \text{tag}^b \}$, where $\text{tag}^a$ is a parent tag, $\text{tag}^b$ is its child tag, and $\text{tag}^a \to \text{tag}^b$ indicates that $\text{tag}^a$ is semantically broader than $\text{tag}^b$.

Since the original textual data describes the objective characteristics of items but rarely reflects subjective user preferences, we use LLMs to \textbf{generate preference summaries} that condense the key features and appeal of each item into a more compact format with higher information related to user preferences. We denote the preference summary for item $i$ as $S_i$.

These steps are accomplished by the following prompt:

\begin{framed}
\noindent \textbf{Item Prompt:}

\noindent Please generate a concise summary highlighting the main features and appeal of the item.

\noindent Please generate 3 level tags for the item. The tags must be different and not duplicated at different levels. Each level can have multiple tags. Higher-level tags should represent broad, general categories that define the item’s overall domain or purpose. Lower-level tags should represent increasingly detailed classifications.

\noindent Please generate edges between different level tags. Each edge should represent a parent-child relationship, where a higher-level tag is the parent and a lower-level tag is the child. The tags on each edge must be within the generated set of tags.

\noindent \textbf{Input:}

\noindent Item Information: $ t_i^1 | t_i^2 | \dots | t_i^m$.

\noindent \textbf{Output:}

\noindent Summary: $\{S_i\}$, 

\noindent Tags: $\{ \text{tag}_{i}^1, \dots, \text{tag}_{i}^{n}\}$, 

\noindent Edges: $\{ \text{tag}^{a_1} \rightarrow \text{tag}^{b_1} , \dots, \text{tag}^{a_k} \rightarrow \text{tag}^{b_k}\}$.
\end{framed}

We explicitly specify the tags' layer as 3 in the prompt. On the one hand, it ensures consistency in the tags generated for different items. On the other hand, three layers of tags allows for a clear distinction between broader and more specific categories.

After generating textual data for items, we use LLMs to \textbf{create user-specific textual data} by analyzing the item summaries and tags associated with user-interacted items. 
This step is critical for capturing the user's preferences in a concise manner. 
We denote the summary of user $u$ as $S_u$ and $u$-interacted items as $\{i_u^1, \dots, i_u^k\}$.

The following prompt is used to generate the user summary:

\begin{framed} 
\noindent \textbf{User Prompt:}

\noindent Please analyze the item summaries and tags and generate a concise summary that captures the user's preferences based on these inputs.
\noindent Ensure that the summary reflects the user's preferences, avoiding repetition while capturing patterns, common themes, and specific features.

\noindent \textbf{Input:}

\noindent Summary: $S_{i_u^1}$, Tags: $\text{Tags}_{i_u^1}$ | \dots | Summary: $S_{i_u^k}$, Tags: $\text{Tags}_{i_u^k}$.

\noindent \textbf{Output:}

\noindent Summary: $\{S_u\}$.
\end{framed}

Due to the hyperbolic model cannot process textual data, we use a text embedding model to \textbf{encode the user and item preference summaries into semantic embeddings}.
Formally, we denote the text embedding model as $\textbf{Encoder}$, which encodes the user and item summaries $S_u$ and $S_i$ into embeddings $\mathbf{s}_u$ and $\mathbf{s}_i$, where:
\begin{equation}
    \begin{aligned}
        \mathbf{s}_u &= \textbf{Encoder}(S_u),
        \quad \quad 
        \mathbf{s}_i = \textbf{Encoder}(S_i).
    \end{aligned}
\end{equation}

\subsection{Meta-optimized Semantic Extraction}
In this module, we extract hierarchical information from semantic embeddings, corresponding to the first phase of the meta-optimized two-phase training strategy. 
Specifically, we employ a MoE model to transform these embeddings from semantic space to hyperbolic collaborative space.
The transformed embeddings are then fed into the hyperbolic model, which is optimized using hyperbolic margin ranking loss.
By freezing both semantic embeddings and hyperbolic model parameters, we leverage the hyperbolic model's ability to capture hierarchical information, enabling the MoE to extract hierarchical information from the semantic embeddings.

Formally, we denote the hyperbolic model as $\textbf{HM}$ and the margin used in this phase as $m_1$. This phase can be represented as:
\begin{equation}
    \begin{aligned}
        \mathbf{x}_u &= \sum_{k=1}^{K} g_k(\mathbf{s}_u^{*}) \cdot \text{Expert}_k(\mathbf{s}_u^{*}),
        \quad
        \mathbf{x}_i  = \sum_{k=1}^{K} g_k(\mathbf{s}_i^{*}) \cdot \text{Expert}_k(\mathbf{s}_i^{*}), 
    \end{aligned}
\end{equation}
\begin{equation}
    \begin{aligned}
        \mathbf{z} &= \mathbf{W}^\text{gate} \mathbf{s} + \mathbf{b}^\text{gate},
        \quad
        g_k(\mathbf{s}) = \frac{\exp(\text{z}_k)}{\sum_{j=1}^{K} \exp(\text{z}_j)},
    \end{aligned}
\end{equation}
\begin{equation}
    \text{Expert}_k(\mathbf{s}) = \mathbf{W}^\text{expert}_k \mathbf{s} + \mathbf{b}^\text{expert}_k,
\end{equation}
\begin{equation}
    \mathcal{L}_{meta} = \textbf{HM}^{*}(\mathbf{x}_u, \mathbf{x}_i, m_1, \mathcal{I}^{user}),
\end{equation}
where $*$ denotes frozen parameters; $K$ is the number of experts; $\text{Expert}_k$ is the $k$-th expert in MoE; $g_k(\cdot)$ is the gating function that computes the weight for $k$-th expert; $\mathbf{W}^\text{gate} \in \mathbb{R}^{d_1 \times K}$ and $\mathbf{W}^\text{expert}_k \in \mathbb{R}^{d_1 \times d_2}$ are weight matrices; $\mathbf{b}^\text{gate} \in \mathbb{R}^{K}$ and $\mathbf{b}^\text{expert}_k \in \mathbb{R}^{d_2}$ are biases; and $\mathcal{L}_{meta}$ is the loss computed by $\textbf{HM}$. 

$\mathcal{L}_{meta}$ comprises the hyperbolic margin ranking loss along with other auxiliary losses of $\textbf{HM}$. The margin $m_1$ defines the minimum separation distance between positive and negative samples in the hyperbolic margin ranking loss of $\mathcal{L}_{meta}$.
We adopt a smaller value for margin $m_1$, enabling the MoE to make subtle adjustments to the semantic embeddings without causing excessive separation, which could hinder effective extraction and lead to overfitting in the subsequent recommender task.

\subsection{Structural and Semantic Integration}
After the Meta-optimized Semantic Extraction module is completed, we \textbf{integrate the extracted structural and semantic hierarchical information into the hyperbolic model}, corresponding to the second phase of the meta-optimized strategy. 
In this phase, we use the MoE-transformed embeddings $\mathbf{x}_u$, $\mathbf{x}_i$ to initialize the user and item ID embeddings, and then train the hyperbolic model.

This phase integrates the \textbf{semantic hierarchical information} into the hyperbolic model and can be represented as:
\begin{equation}
    \begin{aligned}
        \mathbf{e}_u &= \mathbf{x}_u, 
        \quad \quad 
        \mathbf{e}_i = \mathbf{x}_i,
    \end{aligned}
\end{equation}
\begin{equation}
    \mathbf{h}_u, \mathbf{h}_i, \mathcal{L}_{ui} = \textbf{HM}(\mathbf{e}_u, \mathbf{e}_i, m_2, \mathcal{I}^{user}),
\end{equation}
where $\mathbf{e}_u$ and $\mathbf{e}_i$ are the user and item ID embeddings; $\mathbf{h}_u$ and $\mathbf{h}_i$ are the final user and item representations learned from $\mathcal{I}^{user}$; and $\mathcal{L}_{ui}$, $m_2$ are the loss and margin used for learning $\mathcal{I}^{user}$.

Furthermore, we construct a tag-item interaction matrix $\mathcal{I}^{tag}$, where the entry $\mathcal{I}^{tag}_{\text{tag}^a-\text{tag}^b}$ indicates the interaction between $\text{tag}^a$ and $\text{tag}^b$, and the entry $\mathcal{I}^{tag}_{\text{tag}-i}$ indicates the interaction between $\text{tag}$ and its associated item $i$.
Each tag has an ID embedding $\textbf{e}_t$, and we use the same hyperbolic model for $\mathcal{I}^{user}$ to learn $\textbf{e}_t$. 
\begin{equation}
    \mathbf{h}_t, \mathbf{h}_i^{\prime}, \mathcal{L}_{tag} = \textbf{HM}(\mathbf{e}_t, \mathbf{e}_i, m_3, \mathcal{I}^{tag}),
\end{equation}
where $\mathbf{h}_t$, $\mathbf{h}_i^{\prime}$ are the final tag and item representations learned from $\mathcal{I}^{tag}$; and $\mathcal{L}_{tag}$, $m_3$ are the loss and margin used for learning $\mathcal{I}^{tag}$.

$\mathcal{L}_{tag}$ is jointly learned with $\mathcal{L}_{ui}$ to optimize the same hyperbolic model. Finally, we apply contrastive learning to align the item representations obtained from both the tag-item interactions and the user-item interactions. It enables the \textbf{structural hierarchical information} in the tags to be sufficiently learned and effectively integrated into the hyperbolic model.

The overall loss function in this phase is defined as:
\begin{equation}
    \mathcal{L} = \mathcal{L}_{ui} + \mathcal{L}_{tag} + w \cdot \mathcal{L}_{cl},
\end{equation}
\begin{equation}
    \mathcal{L}_{cl}  = \sum_{i \in \mathcal{I}}-\log \frac
                    {\exp\left(-d^2 \left( \mathbf{h}_i, \mathbf{h}_i^{\prime} \right) / \tau\right)}
                    {\sum_{j \in \mathcal{I}} \exp \left(-d^2 \left( \mathbf{h}_i, \mathbf{h}_j^{\prime}\right) / \tau\right)},
\end{equation}
where $\mathcal{L}_{cl}$ is the contrastive learning loss; $\mathcal{I}$ denotes the set of all items; $w$ and $\tau$ are hyper-parameters controlling the weight and temperature of $\mathcal{L}_{cl}$; $d^2$ is the squared hyperbolic distance.

\section{Experiment}
In this section, we perform a series of comprehensive experiments and in-depth analyses to investigate the following questions:
\begin{itemize} [leftmargin=*]
    \item Q1: What is the impact of structural and semantic hierarchical information on the performance of recommender systems?
    \item Q2: What is the contribution of the individual components of HyperLLM to its overall performance?
    \item Q3: Can the meta-optimized strategy accelerate the convergence speed of the subsequent recommender task?
    \item Q4: Is HyperLLM superior to existing LLMs-based recommender frameworks in terms of performance and efficiency?
    \item Q5: What is the impact of HyperLLM on users with different sparsities, and can it address the long-tail effect?
    \item Q6: Can HyperLLM's representations confirm that it learns structural and semantic hierarchical information?
\end{itemize}

\subsection{Experimental Setup}
\begin{table}[h]
    \vspace{-0.15in}
    \captionsetup{skip=0pt}
    \caption{Dataset statistics. Avg Token denotes the average number of tokens in the text Features.}
    \label{table:dataset}
    \centering
    \resizebox{0.5\textwidth}{!}{%
    \begin{tabular}{ccccccc}
    \toprule
    \textbf{Dataset} & \textbf{Users} & \textbf{Items} & \textbf{User-Item} & \textbf{Density} & \textbf{Features} & \textbf{Avg Token} \\
    \midrule
    \textbf{Toys} & 11,268 & 7,309 & 95,420 & 0.12\% & 27,690 & 125  \\
    \textbf{Sports} & 22,686 & 12,301 & 185,718 & 0.07\% & 43,743 & 159  \\
    \textbf{Beauty} & 10,553 & 6,086 & 94,148 & 0.15\% & 22,602 & 138  \\
    \bottomrule
    \end{tabular}
    }
    \vspace{-0.2in}
\end{table}

\begin{table}[h]
    \captionsetup{skip=0pt}
    \caption{Dataset statistics after LLMs-based structural extraction. Avg Summary denotes the average number of tokens in the preference summaries.}
    \label{table:dataset1}
    \centering
    \resizebox{0.42\textwidth}{!}{%
    \begin{tabular}{ccccc}
    \toprule
    \textbf{Dataset} & \textbf{Tags} & \textbf{Tag-Item} & \textbf{Tag-Tag} & \textbf{Avg Summary} \\
    \midrule
    \textbf{Toys} & 10,802 & 25,369 & 61,610 & 54  \\
    \textbf{Sports} & 16,841 & 38,197 & 98,339 & 51 \\
    \textbf{Beauty} & 6,418 & 16,739 & 51,396 & 55 \\
    \bottomrule
    \end{tabular}
   }
   \vspace{-0.2in}
\end{table}

\begin{table*}[h]
  \vspace{-0.15in}
  \centering
  \captionsetup{skip=0pt}
  \caption{Recall(R) and NDCG(N) results for all baselines, baselines with \textbf{Structural} hierarchical information, baselines with \textbf{Semantic} hierarchical information, and baselines with \textbf{HyperLLM}. The best variant for each metric is highlighted in bold.}
  \label{table:performance}
  \resizebox{\textwidth}{!}{
    \begin{tabular}{c|c|cccc|cccc|cccc}
      \toprule
            \multicolumn{2}{c|}{Dataset} & \multicolumn{4}{c|}{Toys} & \multicolumn{4}{c|}{Sports} & \multicolumn{4}{c}{Beauty} \\
        \midrule
            Backbone & Variants & R@10 & R@20 & N@10 & N@20 & R@10 & R@20 & N@10 & N@20 & R@10 & R@20 & N@10 & N@20\\
        \midrule
            \multirow{5}{*}{HGCF}
            & Base & 0.0865 & 0.1211 & 0.0489 & 0.0578 & 0.0581 & 0.0852 & 0.0308 & 0.0377 & 0.1150 & 0.1604 & 0.0651 & 0.0771 \\
            & Structural & 0.0964 & 0.1367 & 0.0533 & 0.0636 & 0.0620 & 0.0944 & 0.0333 & 0.0415 & 0.1176 & 0.1656 & 0.0675 & 0.0801 \\
            & Semantic & 0.1138 & 0.1599 & 0.0628 & 0.0746 & 0.0731 & 0.1091 & 0.0386 & 0.0477 & 0.1281 & 0.1848 & 0.0722 & 0.0870 \\
            & HyperLLM & \textbf{0.1168} & \textbf{0.1638} & \textbf{0.0639} & \textbf{0.0759} & \textbf{0.0754} & \textbf{0.1099} & \textbf{0.0394} & \textbf{0.0482} & \textbf{0.1289} & \textbf{0.1889} & \textbf{0.0736} & \textbf{0.0893} \\ 
            & \textbf{Best Imprv.} & 35.03\% & 35.26\% & 30.67\% & 31.31\% & 29.78\% & 28.99\% & 27.92\% & 27.85\% & 12.09\% & 17.77\% & 13.06\% & 15.82\% \\
        \midrule
            \multirow{5}{*}{HRCF}
            & Base & 0.0867 & 0.1200 & 0.0491 & 0.0576 & 0.0576 & 0.0854 & 0.0308 & 0.0378 & 0.1154 & 0.1607 & 0.0649 & 0.0768 \\
            & Structural & 0.0960 & 0.1374 & 0.0526 & 0.0632 & 0.0613 & 0.0952 & 0.0326 & 0.0412 & 0.1167 & 0.1694 & 0.0666 & 0.0803 \\
            & Semantic & 0.1174 & 0.1654 & 0.0654 & 0.0776 & 0.0750 & 0.1111 & 0.0395 & 0.0487 & \textbf{0.1310} & 0.1890 & 0.0733 & 0.0885 \\
            & HyperLLM & \textbf{0.1208} & \textbf{0.1700} & \textbf{0.0664} & \textbf{0.0789} & \textbf{0.0771} & \textbf{0.1133} & \textbf{0.0403} & \textbf{0.0495} & 0.1297 & \textbf{0.1904} & \textbf{0.0736} & \textbf{0.0895} \\ 
            & \textbf{Best Imprv.} & 39.33\% & 41.67\% & 35.23\% & 36.98\% & 33.85\% & 32.67\% & 30.84\% & 30.95\% & 13.52\% & 18.48\% & 13.41\% & 16.54\% \\
        \midrule
            \multirow{5}{*}{HICF}
            & Base & 0.0945 & 0.1345 & 0.0534 & 0.0636 & 0.0621 & 0.0912 & 0.0338 & 0.0412 & 0.1207 & 0.1685 & 0.0696 & 0.0820 \\
            & Structural & 0.1003 & 0.1378 & 0.0561 & 0.0656 & 0.0655 & 0.0970 & 0.0355 & 0.0435& 0.1239 & 0.1742 & 0.0701 & 0.0833 \\
            & Semantic & 0.1194 & 0.1676 & 0.0666 & 0.0790 & 0.0782 & 0.1168 & 0.0421 & 0.0519 & 0.1389 & 0.1959 & 0.0803 & 0.0953 \\
            & HyperLLM & \textbf{0.1222} & \textbf{0.1690} & \textbf{0.0679} & \textbf{0.0798} & \textbf{0.0792} & \textbf{0.1175} & \textbf{0.0425} & \textbf{0.0523} & \textbf{0.1404} & \textbf{0.1978} & \textbf{0.0807} & \textbf{0.0956} \\ 
            & \textbf{Best Imprv.} & 29.31\% & 25.65\% & 27.15\% & 25.47\% & 27.54\% & 28.84\% & 25.74\% & 26.94\% & 16.32\% & 17.39\% & 15.95\% & 16.59\% \\
        \midrule
           \multirow{5}{*}{HGCH}
            & Base & 0.0989 & 0.1393 & 0.0541 & 0.0645 & 0.0569 & 0.0885 & 0.0299 & 0.0379 & 0.1200 & 0.1727 & 0.0663 & 0.0801 \\
            & Structural & 0.1003 & 0.1424 & 0.0544 & 0.0651 & 0.0599 & 0.0936 & 0.0316 & 0.0402 & 0.1240 & 0.1749 & 0.0699 & 0.0832 \\
            & Semantic & 0.1070 & 0.1536 & 0.0590 & 0.0709 & 0.0702 & 0.1057 & 0.0360 & 0.0451 & 0.1259 & 0.1844 & 0.0711 & 0.0864 \\
            & HyperLLM & \textbf{0.1100} & \textbf{0.1547} & \textbf{0.0604} & \textbf{0.0718} & \textbf{0.0715} & \textbf{0.1065} & \textbf{0.0368} & \textbf{0.0457} & \textbf{0.1296} & \textbf{0.1861} & \textbf{0.0725} & \textbf{0.0872} \\ 
            & \textbf{Best Imprv.} & 11.22\% & 11.06\% & 11.65\% & 11.32\% & 25.66\% & 20.34\% & 23.08\% & 20.58\% & 8.00\% & 7.76\% & 9.35\% & 8.86\% \\
        \midrule
            \multirow{5}{*}{HyperCL}
            & Base & 0.0984 & 0.1381 & 0.0549 & 0.0651 & 0.0641 & 0.0965 & 0.0348 & 0.0431 & 0.1144 & 0.1659 & 0.0637 & 0.0772 \\
            & Structural & 0.0986 & 0.1405 & 0.0550 & 0.0658 & 0.0646 & 0.0980 & 0.0350 & 0.0435 & 0.1165 & 0.1667 & 0.0654 & 0.0785 \\
            & Semantic & 0.1110 & 0.1571 & 0.0611 & 0.0729 & 0.0717 & 0.1096 & 0.0387 & 0.0484 & 0.1207 & 0.1782 & 0.0671 & 0.0820 \\
            & HyperLLM & \textbf{0.1127} & \textbf{0.1599} & \textbf{0.0626} & \textbf{0.0746} & \textbf{0.0724} & \textbf{0.1102} & \textbf{0.0391} & \textbf{0.0486} & \textbf{0.1227} & \textbf{0.1803} & \textbf{0.0677} & \textbf{0.0827} \\ 
            & \textbf{Best Imprv.} & 14.53\% & 15.79\% & 14.03\% & 14.59\% & 12.95\% & 14.20\% & 12.36\% & 12.76\% & 7.26\% & 8.68\% & 6.28\% & 7.12\% \\
      \bottomrule
      \end{tabular}
    }
  \label{tab:overall comparison}%
  \vspace{-0.2in}
\end{table*}%

\subsubsection{Datasets.} 
In the experiments, we use three datasets that are frequently employed in LLMs-based recommender systems: Amazon-Toys, Amazon-Sports, and Amazon-Beauty. We select user-item interactions with ratings $\ge$ 4 and apply 5-core settings to represent implicit feedback preferences. The statistics for these datasets and after LLMs-based structural extraction are shown in Table \ref{table:dataset} and \ref{table:dataset1}, respectively. It can be observed that the preference summaries contain fewer tokens, indicating higher information density. The interactions in each dataset are randomly split into training, validation, and test sets with a ratio of 8:1:1. We evaluate the model's performance on these datasets using Recall@$K$ and NDCG@$K$. To ensure consistency, the model is trained using the best-performing hyper-parameters based on the validation set.

\subsubsection{Baselines.} 
We build HyperLLM upon the following hyperbolic space recommender systems. These baselines include both the most popular methods and the latest advancements. Additionally, these methods utilize various hyperbolic mathematical models and hyperbolic neural network architectures.

\begin{itemize} [leftmargin=*]
\item \textbf{HGCF} \cite{sun2021hgcf} proposes an HGCN based on the Hyperboloid model and optimizes it using margin ranking loss.
\item \textbf{HRCF} \cite{yang2022hrcf} introduces root alignment and geometry-aware regularization loss to HGCF.
\item \textbf{HICF} \cite{yang2022hicf} improves the margin ranking loss with geometry-aware adaptive margins and informative negative sampling.
\item \textbf{HGCH} \cite{zhang2024hgch} implements a fully HGCN on the Poincaré model using the gyromidpoint technique and margin ranking loss.
\item \textbf{HyperCL} \cite{qin2024hyperbolic} designs a fully HGCN on the Hyperboloid model and aligns HGCN with fully HGCN using contrastive learning.
\end{itemize}

\subsubsection{Settings.} We utilize the LLaMA3-8B ~\cite{dubey2024llama} as LLMs to generate textual data and employ the text-embedding-3-large model ~\cite{neelakantan2022text} for encoding preference summaries into semantic embeddings. The dimension $d_1$ of semantic embeddings is 3072 and the dimension $d_2$ of ID embeddings is 50. The number of experts $K$ in MoE is selected from $\{12,16,20\}$. The margin $m_1$ used in meta-optimized phase is selected from $\{0.1,0.2,0.3,0.4,0.5\}$. In most cases, the margin $m_2$ is selected from $[0.1, 4.0]$ and $m_3$ is selected from $[0.1, 1.0]$. The $w$ and $\tau$ in contrastive learning are set to 0.01 and 0.5.
The training epochs is set to 500, with early stopping applied after 50 epochs.
Training-related hyper-parameters, such as the learning rate, are maintained consistent across the same baseline to ensure fairness. Other model-related hyper-parameters are adjusted using grid search. All of our experiments can be completed on a single RTX 3090.

\begin{figure*}[t]
    \vspace{-0.2in}
    \centering
    \captionsetup{skip=0pt}
    \includegraphics[width=1.0\textwidth]{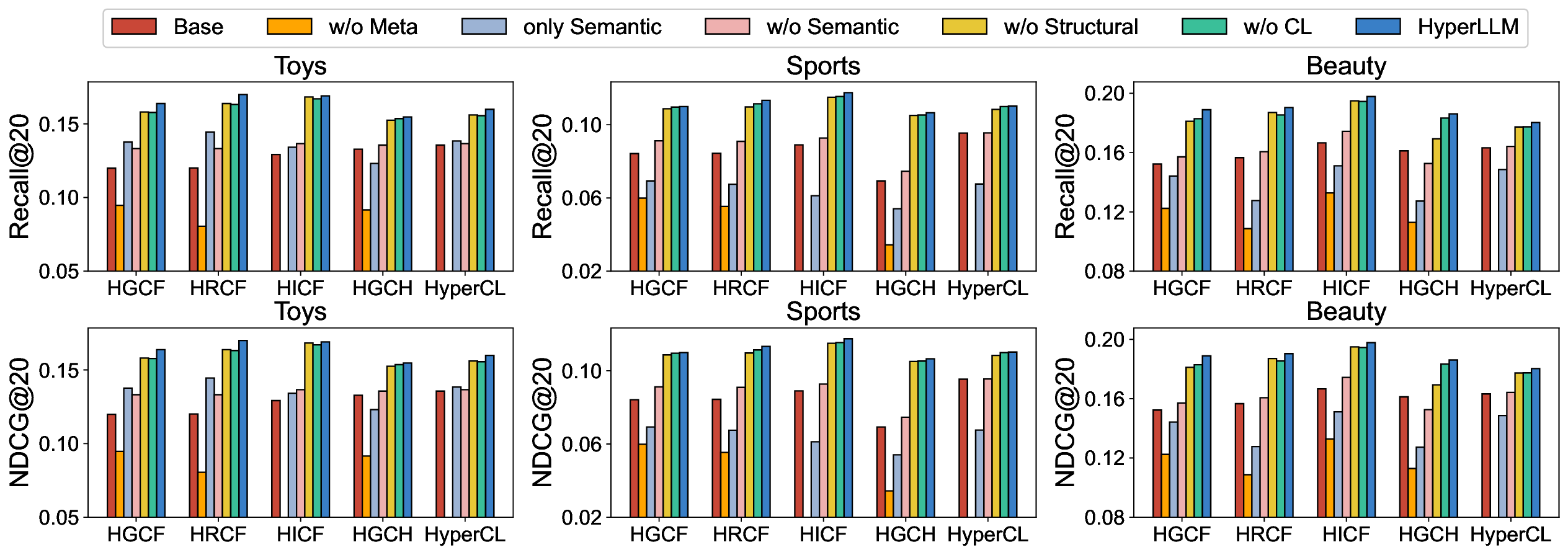}
    \caption{The ablation results for HyperLLM, with missing values for the w/o Meta variant in HICF and HyperCL due to NaN occurrences during the first epoch of training.}
    \label{fig:ablation}
    \vspace{-0.2in}
\end{figure*}

\subsection{Recommender Performance (Q1)}
Table \ref{table:performance} presents the performance of baselines integrated with structural and semantic hierarchical information. 
The \textbf{Structural}, \textbf{Semantic}, and \textbf{HyperLLM} variants represent baselines integrated with structural, semantic, and both structural and semantic hierarchical information, respectively.
The results show that incorporating structural or semantic hierarchical information consistently improves the performance of baselines. Notably, the performance gains from semantic hierarchical information are more substantial, suggesting that the nuanced latent information are more beneficial for hyperbolic space recommender systems. 
HyperLLM leverages both structural and semantic hierarchical information, achieving superior performance compared to the Structural and Semantic variants.
Additionally, HyperLLM achieves maximum improvements of 35.3\%, 41.7\%, 29.3\%, 25.7\%, 15.8\% over HGCF, HRCF, HICF, HGCH, HyperCL, respectively. In terms of average improvements, HyperLLM outperforms these baselines by 25.5\%, 28.5\%, 23.6\%, 14.1\%, 11.7\%, respectively.
It highlights the complementary relationship between these two types of information, each contributing uniquely to the model’s understanding of the semantic hierarchy.

However, it is noteworthy that HyperLLM exhibits relatively smaller improvements when applied to baselines employing fully hyperbolic architectures. Since fully hyperbolic baselines perform all neural operations within hyperbolic space, they are able to more effectively capture  the hierarchical structures inherent in the interaction data. As a result, these baselines are already adept at leveraging hyperbolic geometry to model complex relationships, leaving less room for additional enhancements. 
Nonetheless, HyperLLM still provides meaningful performance gains, highlighting its ability to enhance fully hyperbolic models by integrating the hierarchical information.
In summary, these results demonstrate that the hierarchical information from semantic and textual data can significantly enhance hyperbolic space recommender systems to capture user preferences.

\subsection{Ablation Study (Q2)}
In this subsection, we remove the components from HyperLLM framework to evaluate their impact to the overall recommender performance.
The \textbf{w/o Meta} variant removes the meta-optimized strategy and co-optimizes semantic embeddings and the MoE model with the Structural and Semantic Integration module. We report the result with the higher value between freezing and not freezing the semantic embeddings for the w/o Meta.
The \textbf{only Semantic}, \textbf{w/o Semantic}, and \textbf{w/o Structural} variants remove the Structural and Semantic Integration module, semantic hierarchical information, and structural hierarchical information from HyperLLM, respectively. The \textbf{w/o CL} variant removes the contrastive learning loss from Structural and Semantic Integration module.

As shown in Figure \ref{fig:ablation}, the results demonstrate that each component contributes positively to the overall performance of HyperLLM. 
Among them, components related to semantic hierarchical information consistently make greater contributions than those related to structural information.
Furthermore, the worst-performing variant w/o Meta cannot produce valid results under the HICF and HyperCL baselines, highlighting the significant gap between the semantic and collaborative spaces. In addition, the only Semantic variant significantly outperforms w/o Meta, indicating that the meta-optimized strategy plays an important role in bridging the gap. Meanwhile, only Semantic underperforms the baselines on the Sports and Beauty datasets, which further validates that the purpose of this strategy is to extract hierarchical information from semantic embeddings, allowing it to be better integrated into the recommender task in the next phase.
These findings validate HyperLLM's design, highlighting the crucial role of semantic hierarchical information and the meta-optimized strategy in bridging the gap between semantic and hyperbolic collaborative spaces.

\begin{figure}[t]
    \centering
    \captionsetup{skip=0pt}
    \includegraphics[width=1.0\columnwidth]{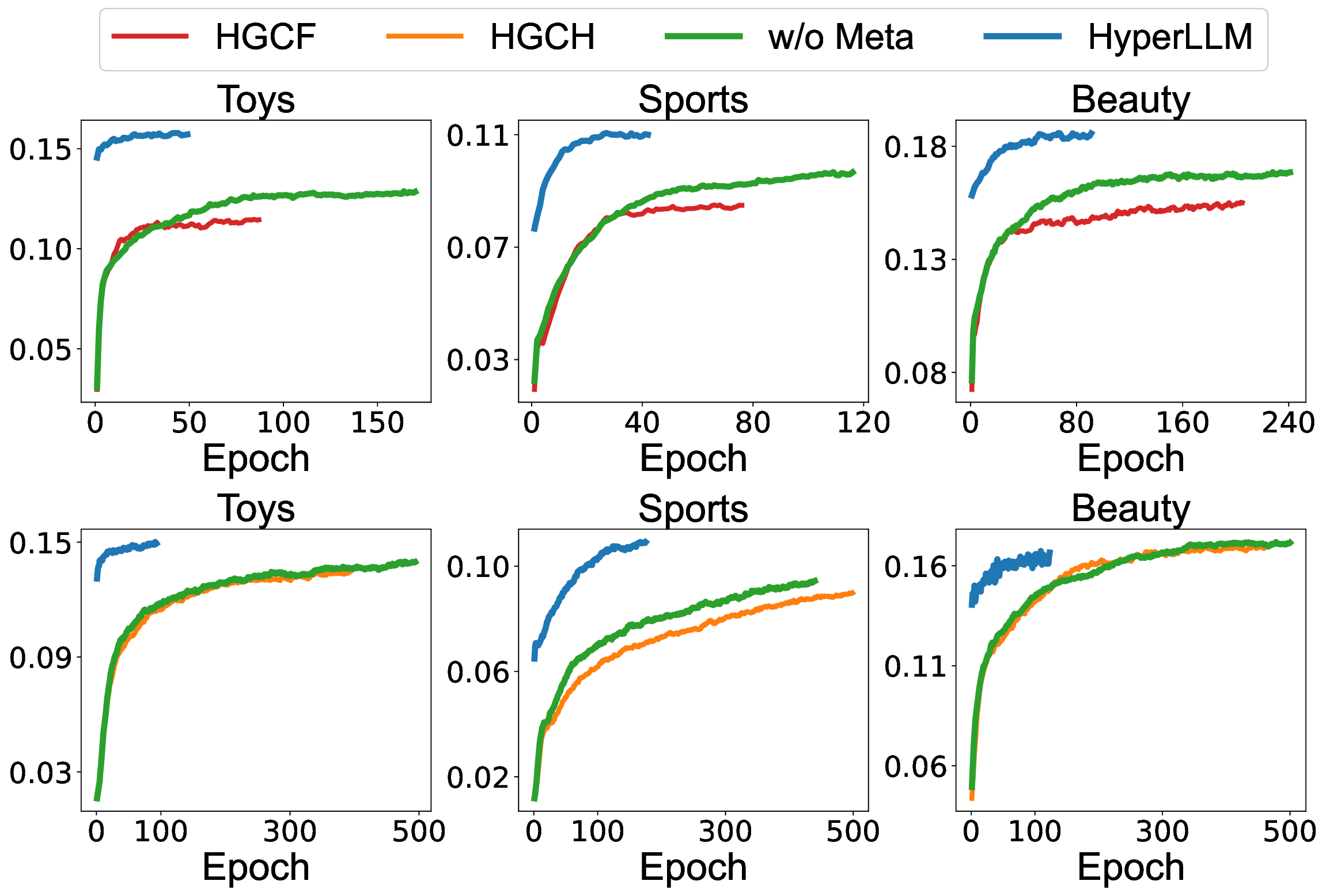}
    \caption{Recall@20 of baselines, HyperLLM without the meta-optimized strategy, and HyperLLM on the validation set at different epochs.}
    \label{fig:meta}
    \vspace{-0.25in}
\end{figure}

\begin{table*}[t]
    \vspace{-0.18in}
    \centering
    \captionsetup{skip=0pt}
    \caption{Recall, NDCG, and Time (Seconds) results for the baselines with different LLMs-based recommender frameworks. The best variant for each metric is highlighted in bold, and the second-best variant is underlined. The percentages indicate the improvement of each variant over the baseline for each metric.}
    \resizebox{\textwidth}{!}{
      \begin{tabular}{c|c|ccc|ccc|ccc}
        \toprule
              \multicolumn{2}{c|}{Dataset} & \multicolumn{3}{c|}{Toys} & \multicolumn{3}{c|}{Sports} & \multicolumn{3}{c}{Beauty} \\
          \midrule
              Backbone & Variants & Recall@20 & NDCG@20 & Time & Recall@20 & NDCG@20 & Time & Recall@20 & NDCG@20 & Time \\
          \midrule
              \multirow{5}{*}{HGCF}
              & Base & 0.1211 & 0.0578 & 15 & 0.0852 & 0.0377 & 34 & 0.1604 & 0.0771 & 39 \\
              & KAR & \underline{0.1331(+9.91\%)} & \underline{0.0620(+7.27\%)} & \textbf{53} & 0.0730(-14.3\%) & 0.0306(-18.8\%) & \underline{724} & 0.1565(-2.43\%) & 0.0715(-7.26\%) & \textbf{15} \\
              & RLMRec-Con & 0.1249(+3.14\%) & 0.0593(+2.60\%) & 1121 & \underline{0.0859(+0.82\%)} & 0.0375(-0.53\%) & 1350 & \underline{0.1619(+0.94\%)} & \underline{0.0773(+0.26\%)} & \underline{219} \\
              & RLMRec-Gen & 0.1205(-0.50\%) & 0.0579(+0.17\%) & 513 & 0.0845(-0.82\%) & \underline{0.0382(+1.33\%)} & 2014 & 0.1610(+0.37\%) & 0.0756(-1.95\%) & 541 \\
              & HyperLLM & \textbf{0.1638(+35.26\%)} & \textbf{0.0759(+31.31\%)} & \underline{261} & \textbf{0.1099(+28.99\%)} & \textbf{0.0482(+27.85\%)} & \textbf{355} & \textbf{0.1889(+17.77\%)} & \textbf{0.0893(+15.82\%)} & 268 \\ 
          \midrule
              \multirow{5}{*}{HRCF}
              & Base & 0.1200 & 0.0576 & 33 & 0.0854 & 0.0378 & 40 & 0.1607 & 0.0768 & 42 \\
              & KAR & 0.1200 (0.00\%) & 0.0545(-5.38\%) & 2028 & 0.0679(-20.5\%) & 0.0282(-25.4\%) & 1088 & 0.1443(-10.2\%) & 0.0638(-16.9\%) & 1071 \\
              & RLMRec-Con & 0.1234(+2.83\%) & \underline{0.0586(+1.74\%)} & \textbf{198} & \underline{0.0863(+1.05\%)} & 0.0380(+0.53\%) & \textbf{232} & 0.1606(-0.06\%) & \underline{0.0772(0.52\%)} & \textbf{274} \\
              & RLMRec-Gen & \underline{0.1255(+4.58\%)} & 0.0585(+1.56\%) & 409 & 0.0854 (0.00\%) & \underline{0.0383(+1.32\%)} & 2202 & \underline{0.1615(+0.50\%)} & 0.0765(-0.39\%) & 645 \\
              & HyperLLM & \textbf{0.1700(+41.67\%)} & \textbf{0.0789(+36.98\%)} & \underline{356} & \textbf{0.1133(+32.67\%)} & \textbf{0.0495(+30.95\%)} & \underline{821} & \textbf{0.1904(+18.48\%)} & \textbf{0.0895(+16.54\%)} & \underline{331} \\ 
          \midrule
              \multirow{5}{*}{HICF}
              & Base & 0.1345 & 0.0636 & 17 & 0.0912 & 0.0412 & 19 & 0.1685 & 0.082 & 49 \\
              & KAR & 0.1397(+3.87\%) & 0.0645(+1.42\%) & \underline{127} & 0.0774(-15.1\%) & 0.0318(-22.8\%) & 336 & 0.1611(-4.39\%) & 0.0761(-7.20\%) & \underline{189} \\
              & RLMRec-Con & \underline{0.1405(+4.46\%)} & \underline{0.0668(+5.03\%)} & \textbf{71} & \underline{0.0943(+3.40\%)} & \underline{0.0420(+1.94\%)} & \textbf{151} & \underline{0.1750(+3.86\%)} & \underline{0.0843(+2.80\%)} & \textbf{73} \\
              & RLMRec-Gen & 0.1303(-3.12\%) & 0.0631(-0.79\%) & 155 & 0.0872(-4.39\%) & 0.0378(-8.25\%) & \underline{222} & 0.1673(-0.71\%) & 0.0808(-1.46\%) & 548 \\
              & HyperLLM & \textbf{0.1690(+25.65\%)} & \textbf{0.0798(+25.47\%)} & 169 & \textbf{0.1175(+28.84\%)} & \textbf{0.0523(+26.94\%)} & 447 & \textbf{0.1978(+17.39\%)} & \textbf{0.0956(+16.59\%)} & 254 \\ 
          \midrule
              \multirow{5}{*}{HGCH}
              & Base & 0.1393 & 0.0645 & 200 & 0.0885 & 0.0379 & 482 & 0.1727 & 0.0801 & 226 \\
              & KAR & 0.0736(-47.2\%) & 0.0350(-45.7\%) & \textbf{109} & 0.0717(-19.0\%) & 0.0316(-16.6\%) & \textbf{141} & 0.1202(-30.4\%) & 0.0550(-31.3\%) & 670 \\
              & RLMRec-Con & 0.1389(-0.29\%) & 0.0643(-0.31\%) & 673 & \underline{0.0869(-1.81\%)} & 0.0367(-3.17\%) & 1927 & \underline{0.1722(-0.29\%)} & \underline{0.0817(+2.00\%)} & \underline{500} \\
              & RLMRec-Gen & \underline{0.1415(+1.58\%)} & \underline{0.0647(+0.31\%)} & 1812 & \underline{0.0869(-1.81\%)} & \underline{0.0368(-2.90\%)} & 4338 & 0.1713(-0.81\%) & 0.0792(-1.12\%) & 1970 \\
              & HyperLLM & \textbf{0.1547(+11.06\%)} & \textbf{0.0718(+11.32\%)} & \underline{195} & \textbf{0.1065(+20.34\%)} & \textbf{0.0457(+20.58\%)} & \underline{785} & \textbf{0.1861(+7.76\%)} & \textbf{0.0872(+8.86\%)} & \textbf{252} \\ 
          \midrule
              \multirow{5}{*}{HyperCL}
              & Base & 0.1381 & 0.0651 & 203 & 0.0965 & 0.0431 & 684 & 0.1659 & 0.0772 & 336 \\
              & KAR & 0.0932(-32.5\%) & 0.0415(-36.3\%) & 5738 & 0.0488(-49.4\%) & 0.0198(-54.1\%) & 9485 & 0.1211(-27.0\%) & 0.0552(-28.5\%) &  1617 \\
              & RLMRec-Con & \underline{0.1392(+0.80\%)} & \underline{0.0651 (0.00\%)} & \underline{1070} & \underline{0.0977(+1.24\%)} & \underline{0.0436(+1.16\%)} & \underline{1344} & \underline{0.1685(+1.57\%)} & \underline{0.0783(+1.42\%)} & \underline{1041} \\
              & RLMRec-Gen & 0.1312(-5.00\%) & 0.0621(-4.61\%) & 2243 & 0.0928(-3.83\%) & 0.0415(-3.71\%) & 7331 & 0.1628(-1.87\%) & 0.0760(-1.55\%) & 3299 \\
              & HyperLLM & \textbf{0.1599(+15.79\%)} & \textbf{0.0746(+14.59\%)} & \textbf{170} & \textbf{0.1102(+14.20\%)} & \textbf{0.0486(+12.76\%)} & \textbf{517} & \textbf{0.1803(+8.68\%)} & \textbf{0.0827(+7.12\%)} & \textbf{246} \\ 
        \bottomrule
        \end{tabular}
      }
    \label{table:framework}%
    \vspace{-0.2in}
\end{table*}%

\subsection{Convergence Analysis (Q3)}
To explore whether the meta-optimized strategy accelerates convergence, we analyze the model's performance across different epochs with and without this strategy, as well as the epochs required for convergence. We choose HGCF and HGCH as baselines for analysis, as they employ different hyperbolic mathematical models and architectures, and the results of other baselines are similar to theirs. 
The \textbf{w/o Meta} variant removes the Meta-optimized Semantic Extraction module from HyperLLM. The w/o Meta and HyperLLM variants are built on the baseline shown in the same figure.

Figure \ref{fig:meta} presents the results of these variants on the validation set at each epoch. The results clearly show that HyperLLM significantly outperform the other two variants at the beginning and also require far fewer epochs to converge. This demonstrates that the meta-optimized strategy can effectively accelerate the convergence speed of recommender task in the next phase.

\subsection{Framework Comparison (Q4)}
To verify the superiority of HyperLLM, we compare it with the following LLMs-based recommender frameworks.
\begin{itemize} [leftmargin=*]
\item \textbf{KAR} \cite{xi2024towards} transforms semantic embeddings using a hybrid MoE model and then concatenates them with ID embeddings to form the final input to the model.
\item \textbf{RLMRec-Con} \cite{ren2024representation} uses contrastive learning to align interaction representations with MLP-transformed semantic embeddings.
\item \textbf{RLMRec-Gen} \cite{ren2024representation} processes the masked embeddings through the model and MLP, and aligns them with semantic embeddings.
\end{itemize}

As illustrated in Table \ref{table:framework}, HyperLLM consistently outperforms all LLMs-based frameworks across all datasets, with a notable performance advantage. This underscores the effectiveness of HyperLLM in utilizing structural and semantic hierarchical information within hyperbolic space—an area where existing frameworks fail to exploit. Furthermore, the performance of compared frameworks varies significantly across different baselines and datasets. 
In many cases, these frameworks deliver only minor improvements or even declines, with KAR in particular experiencing pronounced drops in performance.
In contrast, HyperLLM achieves consistently superior results across all evaluated scenarios, showcasing its robustness and adaptability to diverse data distributions.

In terms of computational efficiency, HyperLLM remains relatively stable across different settings. While it may not always achieve the fastest convergence, its training time stays within an acceptable range. In contrast, the training times of other frameworks fluctuate significantly due to different convergence rates, with some cases requiring more epochs to converge, resulting in higher computational overhead and inefficiency. HyperLLM's ability to effectively extract and integrate hierarchical information facilitates faster convergence, ensuring competitive and stable efficiency.
The comprehensive comparison demonstrates that HyperLLM not only surpasses existing LLMs-based frameworks in terms of performance but also offers more stable training efficiency. 

\begin{figure*}[t]
    \vspace{-0.1in}
    \centering
    \captionsetup{skip=0pt}
    \includegraphics[width=0.96\textwidth]{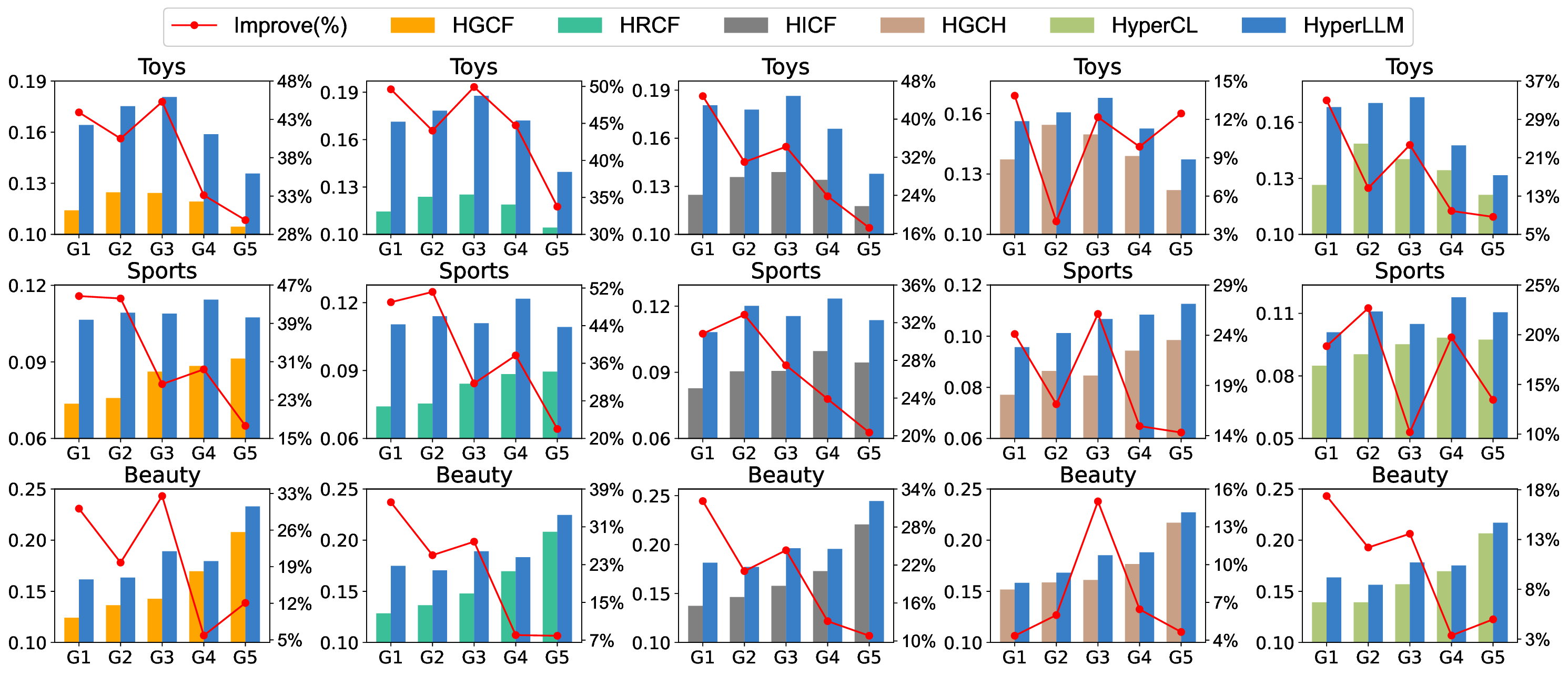}
    \caption{The Recall@20 results for baselines and baselines with HyperLLM on user groups with different levels of sparsity. G1 to G5 represent user groups 1 to 5, with smaller numbers indicating higher sparsity. The right y-axis represents the percentage improvement of HyperLLM compared to the baseline.}
    \label{fig:longtail}
    \vspace{-0.2in}
\end{figure*}

\subsection{Long-Tail Study (Q5)}
To investigate the impact of HyperLLM on users with varying levels of sparsity, we divide users in each dataset into five equally sized groups based on the number of interactions, ordered from least to most. These groups represent different levels of sparsity and help evaluate whether HyperLLM addresses the long-tail effect.

As shown in Figure \ref{fig:longtail}, HyperLLM improves the recommender performance of baselines across all groups, whereas the baselines perform worse in the sparse groups. In contrast, HyperLLM's performance remains relatively stable across different groups and does not exhibit a decreasing trend as the groups become sparser, except in the Beauty dataset. Despite the decreasing trend observed in the Beauty dataset, the performance improvements in the sparse groups are much larger than those in the non-sparse groups. Therefore, this anomaly may be attributed to the inherent limitations of hyperbolic space recommender systems for this dataset. Additionally, HyperLLM's performance improvements increase as the groups become sparser in most cases. The substantial improvements in sparse groups highlight HyperLLM's ability to enhance recommendation quality for users with limited interactions.

This study demonstrates the presence of the long-tail effect and further emphasizes HyperLLM's ability to better address this issue, improving recommender performance where the baselines struggle.
By effectively addressing the long-tail problem, HyperLLM ensures more balanced recommendations, catering to the entire user base.

\subsection{Representation Analysis (Q6)}
To validate that HyperLLM learns structural and semantic hierarchical information, we visualize the learned representations via hierarchical clustering. Specifically, we compute the linkage matrices ~\cite{mullner2011modern} of the representations using \textbf{the ward variance minimization algorithm} to construct dendrograms that reveal their underlying hierarchical structures. This method iteratively merges clusters based on proximity while minimizing variance within each group to optimize cohesion. Consequently, the dendrograms provide visualizations of how representations are hierarchically grouped based on their latent similarities.

\begin{figure}[t]
\centering
\captionsetup{skip=0pt}
\begin{minipage}{0.45\textwidth}
\includegraphics[width=\textwidth]{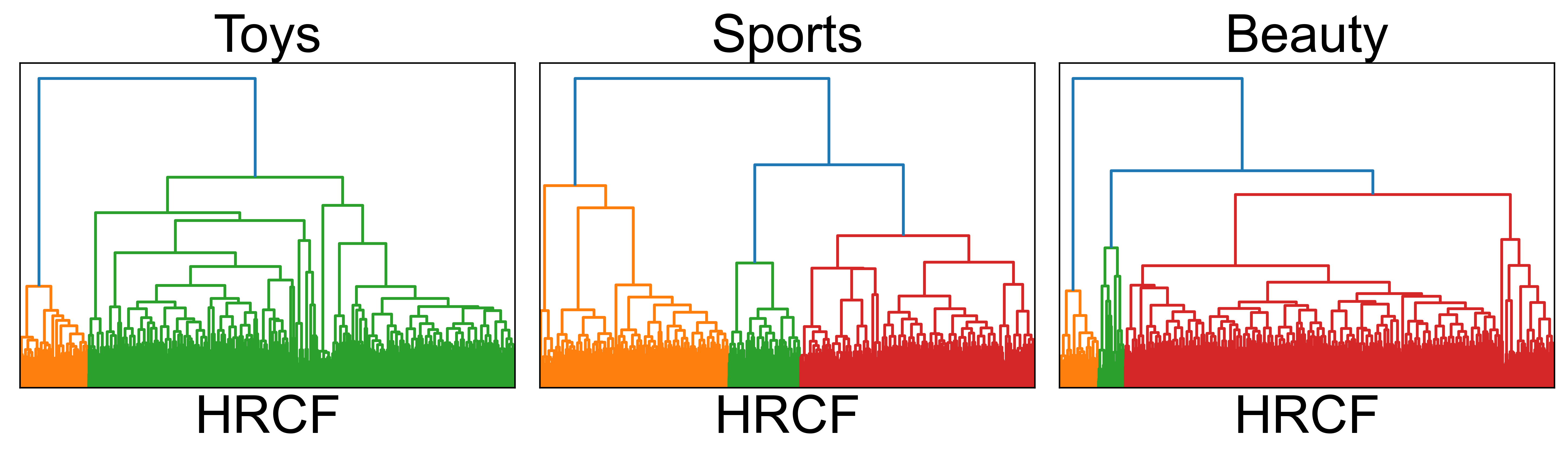}%
\end{minipage}
\hfill
\begin{minipage}{0.45\textwidth}
\includegraphics[width=\textwidth]{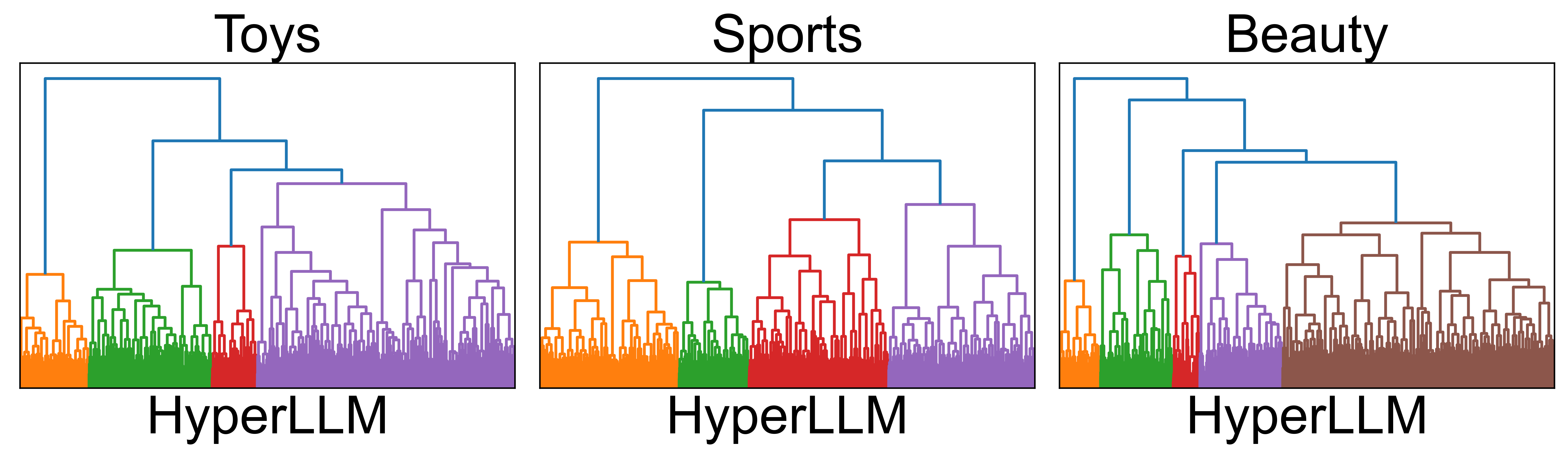}%
\end{minipage}
\caption{Hierarchical clustering visualization of representations for HRCF and HRCF with HyperLLM.}
\label{fig:representation}
\end{figure}

Figure \ref{fig:representation} presents the dendrograms of representations learned by HyperLLM compared to the baseline HRCF. In these dendrograms, different colors represent distinct clusters derived from the hierarchical structures. HyperLLM's dendrograms exhibit a more diverse and clearly separated color distribution compared to the baseline. Additionally, the branches in HyperLLM's dendrograms are more balanced and show deeper hierarchical levels, suggesting a richer modeling of hierarchies. In contrast, the baseline's clusters are less distinct, and the hierarchical separations are less pronounced.
This indicates that the hierarchical clustering for HyperLLM organizes its representations into clusters with more hierarchical structures.
These observations confirm that HyperLLM effectively learns the structural and semantic hierarchical information. 

\section{Conclusion}
In this paper, we propose a model-agnostic framework, named HyperLLM, which extracts structural and semantic hierarchical information from textual recommender data and effectively integrates it into hyperbolic space recommender systems. 
Extensive experiments demonstrate that HyperLLM significantly enhances recommender performance, achieves competitive and stable efficiency, and addresses the long-tail effect.
Furthermore, the meta-optimized two-phase training strategy in HyperLLM offers a novel perspective for future research on semantic fusion in both Euclidean and hyperbolic space recommender systems.

\bibliographystyle{ACM-Reference-Format}
\balance
\bibliography{reference}
\end{document}